\documentclass[a4paper,UKenglish,cleveref, autoref, thm-restate]{lipics-v2021}

\title{Optimizing Safe Flow Decompositions in DAGs}

\titlerunning{Optimizing Safe Flows in DAGs} 

\author{Shahbaz Khan}{Department of Computer Science and Engineering, Indian Institute of Technology Roorkee, India}{shahbaz.khan@cs.iitr.ac.in}{https://orcid.org/0000-0001-9352-0088}{}
                
\author{Alexandru I. Tomescu}{Department of Computer Science, University of Helsinki, Finland}{alexandru.tomescu@helsinki.fi}{https://orcid.org/0000-0002-5747-8350}{}

\authorrunning{S. Khan and A.\,I. Tomescu} 

\Copyright{Shahbaz Khan and Alexandru I. Tomescu} 



\category{} 

\relatedversion{A full version of the paper is available at \url{https://arxiv.org/abs/2102.06480}} 

\supplement{}

\funding{This work was partially funded by the European Research Council (ERC) under the European Union's Horizon 2020 research and innovation programme (grant agreement No.~851093, SAFEBIO) and partially by the Academy of Finland (grants No.~322595, 328877).}

\acknowledgements{We would like to thank Manuel C\'{a}ceres from University of Helsinki, for helpful discussions and for highlighting several errors in our previous approaches.}

\nolinenumbers 

\hideLIPIcs  

\EventEditors{Shiri Chechik, Gonzalo Navarro, Eva Rotenberg, and Grzegorz Herman}
\EventNoEds{4}
\EventLongTitle{30th Annual European Symposium on Algorithms (ESA 2022)}
\EventShortTitle{ESA 2022}
\EventAcronym{ESA}
\EventYear{2022}
\EventDate{September 5--9, 2022}
\EventLocation{Berlin/Potsdam, Germany}
\EventLogo{}
\SeriesVolume{244}
\ArticleNo{88}

\usepackage{amsmath}
\usepackage{amssymb}
\usepackage{hyperref}
\usepackage[dvipsnames]{xcolor}
\usepackage[ruled,vlined]{algorithm2e}  
\usepackage{multicol}
\usepackage{multirow}
\usepackage{comment}

\usepackage{caption}
\usepackage{subcaption}

\newcommand{\sbzk}[1]{\textcolor{cyan}{}}

\begin{document}

\maketitle

\begin{abstract}
Network flow is one of the most studied combinatorial optimization problems having innumerable applications. Any flow on a directed acyclic graph $G$ having $n$ vertices and $m$ edges can be decomposed into a set of $O(m)$ paths. The applications of such a flow decomposition range from network routing to the assembly of biological sequences. However, in some applications, each solution (decomposition) corresponds to some particular data that generated the original flow. Given the possibility of multiple optimal solutions, no optimization criterion ensures the identification of the correct decomposition. Hence, recently flow decomposition was studied [RECOMB22] in the Safe and Complete framework, particularly for RNA Assembly. The proposed solution reported {\em all} the {\em safe} paths, i.e., the paths which are subpath of every possible solution of flow decomposition. 

They presented a characterization of the safe paths, resulting in an $O(mn+out_R)$ time algorithm to compute all safe paths, where $out_R$ is the size of the raw output reporting each safe path explicitly. They also showed that $out_R$ can be $\Omega(mn^2)$ in the worst case but $O(m)$ in the best case. Hence, they further presented an algorithm to report a concise representation of the output $out_C$ in $O(mn+out_C)$ time, where $out_C$  can be $\Omega(mn)$ in the worst case but $O(m)$ in the best case. 

In this work, we study how different safe paths interact, resulting in  optimal output-sensitive algorithms requiring $O(m+out_R)$ and $O(m+out_C)$ time for computing the existing representations of the safe paths. Our algorithm uses a novel data structure called {\em Path Tries}, which may be of independent interest. Further, we propose a new characterization of the safe paths resulting in the {\em optimal} representation of safe paths $out_O$, which can be $\Omega(mn)$ in the worst case but requires optimal $O(1)$ space for every safe path reported. We also present a near-optimal algorithm to compute all the safe paths in $O(m+out_O\log n)$ time. The new representation also establishes tighter worst case bounds $\Theta(mn^2)$ and $\Theta(mn)$ bounds for $out_R$ and $out_C$ (along with $out_O$), respectively.  

Overall we further develop the theory of safe and complete solutions for the flow decomposition problem, giving an optimal algorithm for the explicit representation, and a near-optimal algorithm for the optimal representation of the safe paths. 
\end{abstract}

\newpage
\section{Introduction}

\setcounter{page}{1}
Network flow is one of the most studied problems in theoretical computer science with innumerable applications. 
For a flow network with a unique source $s$ and a unique sink $t$, every valid flow can be decomposed into a set of weighted $s$-$t$ paths and cycles~\cite{FordF10}. For a directed acyclic graph (DAG) such a decomposition contains only paths. Such path (and cycle) view of a flow indicates how information optimally passes from $s$ to $t$, being a key step in network routing problems (e.g.~\cite{hong2013achieving,cohen2014effect,hartman2012split,mumey2015parity}), transportation problems~(e.g.~\cite{Ohst:2015aa,Olsen:2020aa}), or in the more recent and prominent application of reconstructing biological sequences~(\emph{RNA transcripts}, see e.g.~\cite{pertea2015stringtie,tomescu2013novel,gatter2019ryuto,bernard2013flipflop,TomescuGPRKM15,williams2019rna}, or \emph{viral quasi-species genomes}, see e.g.~\cite{DBLP:conf/recomb/BaaijensSS20,BaaijensRKSS19}).

Finding the minimum flow decomposition (i.e., having the minimum number of paths and cycles) is NP-hard, even if the flow network is a DAG~\cite{vatinlen2008simple}. This hardness result led to research on approximation algorithms~\cite{hartman2012split,SUPPAKITPAISARN2016367,pienkosz2015integral,mumey2015parity,baier2005k}, and FPT algorithms~\cite{kloster2018practical}. Practical approaches usually employ the standard \emph{greedy width} heuristic~\cite{vatinlen2008simple}, repeatedly removing an $s$-$t$ path carrying the most amount of flow. Recently, another pseudo-polynomial-time heuristic was proposed~\cite{shao2017theory} for biological data, which tries to iteratively simplify the graph such that the flow decomposition problem can be solved locally at some vertices.

In the routing and transportation applications, an optimal flow decomposition indicates how to send some information from $s$ to $t$, and thus any optimal decomposition is satisfactory. 
However, this is not the case in the prominent application of reconstructing biological sequences, since each flow path represents a reconstructed sequence: a different optimal set of flow paths encodes different biological sequences, which may differ from the real ones. For a concrete example, consider the following application. 
In complex organisms, a gene may produce more RNA molecules (\emph{RNA transcripts}, i.e., strings over an alphabet of four characters), each having a different abundance. Currently, given a sample, one can read the RNA transcripts and find their abundances using \emph{high-throughput sequencing}~\cite{citeulike:3614773}. This technology produces short overlapping substrings of the RNA transcripts. The main approach for recovering the RNA transcripts from such data is to build an edge-weighted DAG from these fragments and to transform the weights into flow values by various optimization criteria, and then to decompose the resulting flow into an ``optimal'' set of weighted paths (i.e., the RNA transcripts and their abundances in the sample)~\cite{DBLP:books/cu/MBCT2015}.  Clearly, if there are multiple optimal flow decomposition solutions, then the reconstructed RNA transcripts may not match the original ones, and thus be incorrect. Thus, the best possible solution is to find \textit{whatever} can be \textit{safely} reported as being \textit{correct}. 

\subsection{Problem Definition and Related Work}
Recently,  Ma et al.~\cite{DBLP:conf/wabi/MaZK20} were the first to address the issue of multiple solutions to the flow decomposition problem, under a probabilistic framework. Later, they~\cite{findingranges} solve a problem (\emph{AND-Quant}), which, in particular, leads to a quadratic-time algorithm for the following problem: given a flow in a DAG, and edges $e_1,e_2,\dots,e_k$, decide if in \emph{every} flow decomposition there is always a decomposed flow path passing through all of $e_1,e_2,\dots,e_k$. Thus, by taking the edges $e_1,e_2,\dots,e_k$ to be the edges of a path $p$, the AND-Quant problem can decide if a path $p$ (i.e., a given biological sequence) appears in all flow decompositions. This indicates that $p$ is likely part of some original RNA transcript.

Another popular approach to address the issue of multiple solutions is the \emph{safety} framework,
 which was introduced by Tomescu and Medvedev~\cite{tomescu2017safe} for the genome assembly problem from bioinformatics. For a problem admitting multiple solutions, a partial solution is said to be \emph{safe} if it appears in all solutions to a problem. For the flow decomposition problem, a path $p$ is safe if for \emph{any} flow decomposition into paths ${\cal P}=\{p_1,\dots,p_k\}$, it holds that $p$ is a subpath of some $p_i$. Considering the weight, a path $p$ is further called \emph{$w$-safe} if, in \textit{any} flow decomposition, $p$ is a subpath of some path(s) in ${\cal P}_f$ whose total weight is at least $w$. 

Khan et al.~\cite{KhanMMLA22} built upon the AND-Quant problem by addressing flow decomposition under the \emph{safety} framework. They presented a local characterization of safe flow paths as compared to the global characterization of AND-Quant. It was directly adaptable to give an optimal verification algorithm, and a simple enumeration algorithm enumerating all safe paths in $O(mn+out)$ time by applying the characterization on a candidate flow decomposition using the standard \textit{two pointer algorithm}\footnote{Along a sample solution, the keeping the left end at the start, the right end is moved along the solution as long as the path is safe (evaluated using verification algorithm). This is reported as a maximal safe. Then right end is extended by an edge making the path unsafe, followed by moving the left pointer right until it is safe again. The process is repeated to report the next maximal safe path, and so on.}. They presented the maximal safe paths $out$ in two formats, the raw output $out_R$ reported each safe path explicitly, and a concise representation $out_C$ which combined the safe paths occurring contiguously in the candidate flow decomposition. Using a worst case example they also proved that the size of $out_R$ can be $\Omega(mn^2)$ in the worst case and $O(m)$ in the best case, whereas that of $out_C$ can be $\Omega(mn)$ in the worst case and $O(m)$ in the best case. However, in their solution the concise representation of the solution depends on the underlying candidate solution used, which hence does not optimize the concise representation. Moreover, they did not address whether the concise representation is the most succinct approach to represent the safe paths.



\subsection{Our results}
Our main contributions can be described as follows:

\begin{enumerate}
    \item \textbf{Merge-Diverge Property of safe paths.} We develop the theory of safe paths for flow decomposition further by studying the conditions for interaction of safe paths. We prove that two safe paths cannot {\em merge} at a vertex (or a set of vertices) and later {\em diverge}. 
    
    \item \textbf{Optimal output-sensitive enumeration algorithms for the current representations.} 
    We use the {\em merge-diverge} property to present optimal output-sensitive algorithms for enumerating all safe paths explicitly in $O(m+out_R)$ time and their optimal concise representation in $O(m+out_C)$ time. Our algorithms uses a novel application of the Trie on paths, referred as {\em Path Tries} which may be of independent interest.

    \item \textbf{Optimal representation of safe paths.} We present a novel characterization of safe paths $out_O$ allowing us to represent a safe path optimally, requiring $O(1)$ space for every reported path. 
    
    \begin{remark}
    In the worst case both concise representation $out_C$~\cite{KhanMMLA22} and our optimal representation $out_O$ may require $\Omega(mn)$ space, however space required per reported path can be much larger for $out_C$ than the optimal $O(1)$ of $out_O$. 
    \end{remark}
    
    \item \textbf{Near optimal algorithm for the optimal output format.} We present an algorithm to report all safe paths using the optimal representation in $O(m+out_O\log n)$ time.
    
    \item \textbf{Tighter worst case bounds on $out_R$ and $out_C$.} Our  characterization allows us to prove matching upper bounds for the worst case lower bounds~\cite{KhanMMLA22} on $out_R$ and $out_C$.
\end{enumerate}

\section{Preliminary}
\label{sec:prelims}
Consider a directed acyclic flow graph $G=(V,E)$ with $|V|=n$ vertices and $|E|=m$ edges, where each edge $e$ has a flow (or weight) $f(e)$ passing through it. For simplicity we assume the graph is connected giving $m\geq n-1$. 
For each vertex $u$, $f_{in}(u)$ and $f_{out}(u)$ denotes the total flow on its incoming edges and total flow on its outgoing edges, respectively. A vertex $v$ in the graph is called a {\em source} if $f_{in}(v)=0$ and a {\em sink} if $f_{out}(v)=0$. The set of sources and sinks of the graph $G$ is denoted by $Source(G)$ and $Sink(G)$ respectively. Every other vertex $v$ satisfies the {\em conservation of flow} $f_{in}(v)=f_{out}(v)$, making the graph a {\em flow graph}. 

For the vertex $u$, $f_{max}(\cdot,u)$ (or $f_{max}(u,\cdot)$) denotes the maximum value of flow on the incoming edges (or outgoing edges) of $u$. 
The corresponding edge is represented by $e_{max}(\cdot,u)$ (or $e_{max}(u,\cdot)$) and its other endpoint (except $u$) is represented by $v_{max}(\cdot,u)$ (or $v_{max}(u,\cdot)$). Note that in case multiple incoming edges (or outgoing edges) have the maximum flow value, we prefer the edge whose other endpoint (except $u$) appears first in the topological order, making $e_{max}(\cdot,u)$ (or $e_{max}(u,\cdot)$) and $v_{max}(\cdot,u)$ (or $v_{max}(u,\cdot)$) distinct. Hence, it is referred as  {\em preferred} maximum incoming (or outgoing) edge/vertex. Further, we represent $e^*_{max}(\cdot,u)$ (or $e^*_{max}(u,\cdot)$) as the {\em unique} maximum incoming (or outgoing) edge if $f_{max}(\cdot,u)$ (or $f_{max}(u,\cdot)$) corresponds to exactly one edge making it equal to  $e_{max}(\cdot,u)$ (or $e_{max}(u,\cdot)$) in such a case, and null otherwise. We similarly define its other endpoint (except $u$) $v^*_{max}(\cdot,u)$ (or $v^*_{max}(u,\cdot)$) which is called {\em unique} maximum incoming (or outgoing) vertex.

For a path $p$ in the graph, $|p|$ represents the number of its edges. A vertex $u$ is called as being on the {\em left} of a vertex $v$ on the path, if $v$ is reachable from $u$ on the path. Similarly, in such a case the vertex $v$ is called as being on the {\em right} of a vertex $u$ on the path. For any path $p$ (or edge) we define its {\em left extension} to be a path created from $p$ by repeatedly prepending the path with the unique maximum incoming edge of the first vertex of the (updated) path. Similarly, we define the {\em right extension} of a path to be a path created by repeatedly adding the unique maximum outgoing edge of the last vertex of the (updated) path.



\begin{figure}[ht]
    \centering
\includegraphics[trim=5 7 10 7, clip,scale=.9]{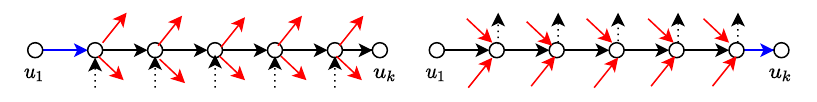}
    \caption{The excess flow of a path is the incoming or outgoing flow  ({\em blue}) that passes through the path despite the flow ({\em red}) leaking at its internal vertices (reproduced from \cite{KhanMMLA22}).}
    \label{fig:excessP}
\end{figure}

The {\em flow decomposition} of $G$ is a set of weighted {\em paths} ${\cal P}_f$ such that the flow on each edge in the $G$ equals the sum of the weights of the paths containing it. A path $p$ is called \emph{$w$-safe} if, in every possible flow decomposition, $p$ is a subpath of some paths in ${\cal P}_f$ whose total weight is at least $w$. A $w$-safe path with $w > 0$, is called a {\em safe flow path}, or simply {\em safe path}. A safe path is {\em left maximal} (or {\em right maximal}) if extending it to the left (or right) with any edge makes it unsafe. A safe path is {\em maximal} if it is both left and right maximal. The safety of a path can be characterized by its \emph{excess flow} (see \Cref{fig:excessP}) and properties of safe paths, described as follows.

\begin{definition}[Excess flow~\cite{KhanMMLA22}]
The \emph{excess flow} $f_p$ of a path $p=\{u_1,u_2,...,u_k\}$ is 
\begin{equation*}
f_p= f(u_1,u_2) - \sum_{\substack{u_i\in \{u_2,...,u_{k-1}\} \\ 
v\neq u_{i+1}}} f(u_i,v)
= f(u_{k-1},u_k) - \sum_{\substack{u_i\in \{u_2,...,u_{k-1}\} \\ 
v\neq u_{i-1}}} f(v,u_i) \end{equation*}
where 
the former and later equations are called {\em diverging} and {\em converging} criterion, respectively.
\label{def:dispConv}
\end{definition}

\begin{restatable}{theorem}{flowSafety}[Safe flow paths~\cite{KhanMMLA22}] Safety of flow  decomposition satisfy the following.
\begin{enumerate}[(a)]
    \item A path $p$ is $w$-safe iff its excess flow $f_p\geq w> 0$.
 \label{prop:flowSafety}
 \item The converging and diverging criteria for a path $p=\{u_1,\cdots,u_k\}$ are equivalent to
 \[f_p = \sum_{i=1}^{k-1} f(u_i,u_{i+1}) - \sum_{i= 2}^{k-1} f_{out}(u_i) =  \sum_{i=1}^{k-1} f(u_i,u_{i+1}) - \sum_{i=2}^{k-1} f_{in}(u_i).\]  
\label{prop:altCriteria}
\item Adding an edge $(u,v)$ to the start or the end of a path in the flow graph, reduces its excess flow by
$f_{in}(v) - f(u,v)$, or $f_{out}(u)-f(u,v)$, respectively.
\label{prop:addEdge}
\end{enumerate}
\label{thm:flowPropO}
\end{restatable}

Additionally, we use the following data structure for answering the level ancestor queries.

\begin{theorem}[Level Ancestors~\cite{BenderF04}]
A given tree with $n$ vertices can be preprocessed in $O(n)$ time to report the level ancestor $LA(v,d)$ for a vertex $v$ at a depth $d$ in $O(1)$ time.
\label{thm:LA}
\end{theorem}

\section{Interaction of Safe Paths}
\label{sec:merge-diverge}
The previous work~\cite{KhanMMLA22} focused on properties of safe paths useful for applying the characterization directly in verification and enumeration algorithms. We now explore further properties of safe walks particularly related to the interaction of safe paths and its consequences. 

\begin{restatable}{lemma}{mergeDiverge}[Merge Diverge]
Two safe paths cannot merge (through distinct edges) at an intermediate vertex (or vertices) and then diverge (through distinct edges).
\label{prop:merge-diverge}
\end{restatable}
\begin{proof}
Let two safe paths $p$ and $p'$ merge at a vertex $v_1$, entering $v_1$ respectively by distinct edges $e_1$ and $e'_1$, and then diverge at a vertex $v_2$, leaving $v_2$ respectively by distinct edges $e_2$ and $e'_2$ (see \Cref{fig:merge-diverge}). 
 
\begin{figure}[ht]
    \centering
\includegraphics[trim=5 10 10 5, clip, scale=1]{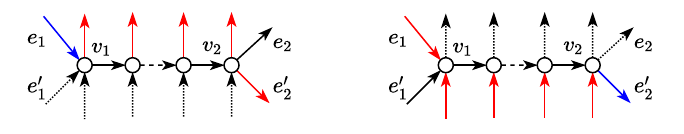}
    \caption{The diverging and converging criterion applied to path $p$ and $p'$ respectively.}
    \label{fig:merge-diverge}
\end{figure}

Using \Cref{thm:flowPropO}\ref{prop:addEdge} we know that removing an edge from the end of a path increases the excess flow and hence remains safe. Thus, the subpaths $\{e_1\cdots,e_2\}$ and $\{e'_1,\cdots,e'_2\}$ of safe paths $p$ and $p'$ respectively, are also safe. By diverging criterion of the safe path $p$ we have $f(e_1)>f(e'_2)$. On the other hand, by converging criterion of the safe path $p'$ we have $f(e'_2)>f(e_1)$, which is a contradiction. 
\end{proof}

This {\em merge-diverge} property has an interesting consequence on the structure of a safe path having an edge that is not a unique maximum outgoing edge of a vertex.

\begin{lemma}
Any safe path having an edge $e(u,v)\neq e^*_{max}(u,\cdot)$, can be extended to the left using only unique maximum incoming edges.
\label{prop:nonMaxOut}
\end{lemma}
\begin{proof}
Consider a path $p$ containing $e_1(u,v)\neq e^*_{max}(u,\cdot)$, which extends to the left of $u$ using edges containing an edge which is not a unique maximum incoming edge $e_2(x,y)\neq e^*_{max}(\cdot,y)$. Now, using \Cref{thm:flowPropO}\ref{prop:addEdge} we know subpath created by removing edges from the end is safe, as removing such edges only increases the excess flow. Hence, the subpath $p:\{e_2\cdots e_1\}$ is safe. Further, \Cref{thm:flowPropO}\ref{prop:addEdge} also implies that a path $p'$ replacing $(x,y)$ with an alternate $e'_2=e_{max}(\cdot,y)$, and $(u,v)$ with an alternate $e'_1=e_{max}(u,\cdot)$ is also safe, as we replace an edge with another having at least the same weight. Note that $e'_1$ and $e'_2$ always exists since $e_1$ and $e_2$ are not unique maximum edges.
Thus, both $p:\{e_2\cdots e_1\}$ and $p':\{e'_2\cdots e'_1\}$ are safe which merge at $y$ and then diverge at $u$ using distinct edges, which is a contradiction.
\end{proof}

\section{Optimal computation of Raw Safe paths}
\label{sec:optSimp}

The essential bottle-neck of the previous approach~\cite{KhanMMLA22} was the use of a candidate flow decomposition, on whose subpaths the safety criteria was evaluated. The computation of a candidate flow decomposition itself requires $O(mn)$ time making it suboptimal. In order to avoid it we are required to process the graph in a structured manner. Given the graph is a DAG, the topological ordering of the graph serves this purpose. 

\begin{figure}
    \centering
    \includegraphics[scale=0.75]{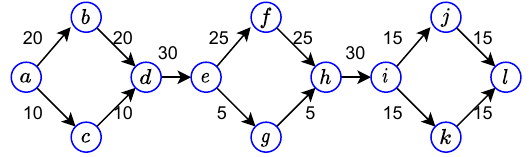}
    \caption{Problem with the simplistic approach. While processing the vertex $e$, we get two safe paths $p_1:<a,b,d,e>$ and $p_2:<a,c,d,e>$. As we continue processing to reach $h$, both $p_1$ and $p_2$ are extended through $f$ to get $p_{11}:<a,b,d,e,f,h,i>$  and $p_{21}:<a,c,d,e,f,h,i>$. However, when extending through $g$ both get trimmed to give the same $p_{12},p_{22}:<d,e,g,h,i>$. Further, the paths $p_{11}$ and $p_{22}$ are again extended through $j$ and $k$ (recall the two pointer algorithm) to get four paths, $p_{111},p_{211}:<d,e,f,h,i,j,l>$ and $p_{112},p_{212}:<d,e,f,h,i,k,l>$. Moreover, paths $p_{12},p_{22}$ can also be extended through $j$ and $k$ to get $p_{121},p_{221}:<h,i,j,l>$  and $p_{122},p_{222}:<h,i,k,l>$. We thus obtain duplicate paths representing the same safe paths from different sources, and some non-maximal paths ($p_{122},p_{222}$) which are subpath of the other reported paths.}
    \label{fig:baseGraph}
\end{figure}

A simple approach is to follow the topological order and maintain all maximal safe paths ending at the currently processed vertex explicitly. And use the two pointer algorithm to extend it as we continue processing the vertices in the topological order. However, to avoid duplicate and non-maximal results we need to identify the common suffixes of the safe paths, which can be processed accordingly (see \Cref{fig:baseGraph}).
Fortunately, for strings the data structure Trie (considered on reversed strings) serves exactly for the same purpose which motivates us to use Tries for storing all the left maximal safe paths ending at a vertex as follows.

\subsection{Data structures ${\cal T}_u$ and ${\cal L}_u$}
We build a Trie structure ${\cal T}_u$ treating the reverse paths ending at a vertex $u$ as strings, such that the common suffixes of the paths are combined. Note that a vertex $v$ can appear multiple times in the Path Trie, if multiple paths containing $v$ do not share $v$ in their common suffix.

\begin{figure}
    \centering
    \includegraphics[scale=0.75,trim={0 .35cm 0 0}, clip]{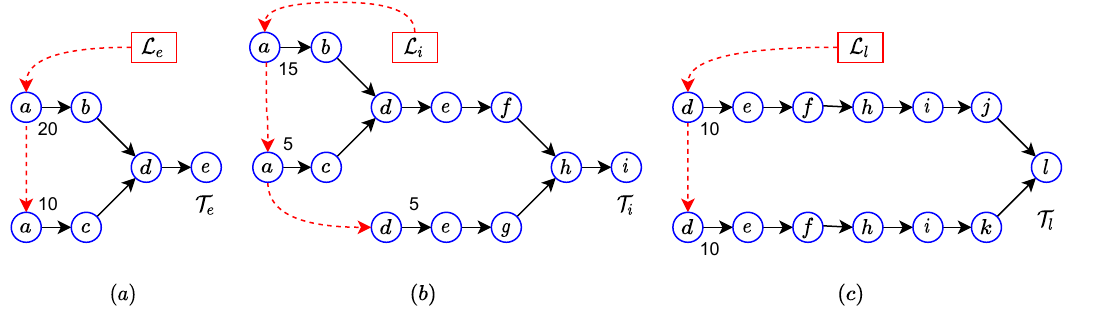}
    \caption{Path Trie structure storing the left maximal paths ending at vertices (a) ${\cal T}_e$ with ${\cal L}_e$, (b) ${\cal T}_i$ and ${\cal L}_i$, and (c)  ${\cal T}_l$ and ${\cal L}_l$.}
    \label{fig:PathTrie}
\end{figure}

All left maximal safe paths ending at a vertex $u$ are hence maintained in ${\cal T}_u$. Additionally, we maintain a linked list ${\cal L}_u$ of the leaves of ${\cal T}_u$ along with the path's corresponding excess flow, i.e. ${\cal L}_u=\{(v_1,f_1),(v_2,f_2),\cdots\}$. Consider \Cref{fig:PathTrie}, we show the path tries at the vertices $e,i$ and $l$ for the graph shown in \Cref{fig:baseGraph}. Note that the leaves represent the left maximal safe paths without repetition or storing subpaths as in the simplistic approach. 
 
\subsection{Algorithm}
The main idea behind our approach is to uniquely extend each safe path ending at a vertex to its preferred maximum out-neighbour in constant time associated with each edge.  For the rest of the out-neighbours, we can build their safe paths from scratch at the expense of the path length. This requires us to process the vertices in the topological order of the graph, such that all the safe paths ending at a vertex are computed before it is processed. We maintain the left maximal safe paths ending at a vertex $u$ in the Path Trie ${\cal T}_u$ and the list of safe paths ${\cal L}_u$. Our algorithm uses optimal $O(m+out_R)$ time, where $out_R$ is the size of the raw output, i.e., each safe path stored explicitly, which is optimal.  

\Cref{alg:optSimp} describes our approach, where vertices are processed in topological order such that while processing a vertex $u$, all the left maximal safe paths ending at $u$ (along with their excess flow) are stored in ${\cal T}_u$ and ${\cal L}_u$. 
While processing $u$ all its safe paths are evaluated for a possible extension to its preferred maximum outgoing neighbour $v^*=v_{max}(u,\cdot)$, which always exists except when $u$ is a sink. Note that $v^*$ is not the unique maximum outgoing neighbour ($v^*_{max}(u,\cdot)$) rather preferred ($v_{max}(u,\cdot)$), so that ${\cal T}_u$ can be used to extend to some vertex in case there is no unique maximum.  Additionally, when $u$ is a source, we have an empty ${\cal T}_u$, so we have no safe paths to extend. Hence, for the other cases we check all paths in ${\cal L}_u$ for possible extension to $v^*$ using \Cref{thm:flowPropO}\ref{prop:addEdge}. 
For this we add the complete ${\cal T}_u$ as a child of ${\cal T}_v$. Thereafter, we need to trim the prefixes of paths in ${\cal T}_{v^*}$ which are not safe and compute the list of safe paths ${\cal L}_{v^*}$. Further, we need to add the safe paths using every other outgoing edge $(u,v)$ ($v\neq v^*$) in the corresponding ${\cal T}_v$. 

We now process each path in ${\cal L}_u$. The paths which are not safe on extending to $(u,v^*)$ are clearly right-maximal (and hence maximal), and hence are reported in the solution $\textsc{Sol}$. We extend all the paths in ${\cal L}_u$ with $(u,v^*)$, and add their maximal suffixes which are safe to ${\cal T}_{v^*}$. Note that the entire path may be safe or at least the edge $(u,v^*)$ is safe. So we start trimming ${\cal T}_{v^*}$ until the path is safe. This is done by maintaining in $f_x$ the excess flow of the path from $x$ to $v^*$, where $f_x$ is updated using \Cref{thm:flowPropO}\ref{prop:addEdge}. When $(u,v^*)$ is added $f_x$ may become negative in which case the path is trimmed from the left until $f_x$ is positive. Now, since in a Trie an edge can be shared by multiple paths having a common suffix, we trim the edge only if it is a leaf, and similarly add to ${\cal L}_{v^*}$ only if it starts from a leaf. Hence, multiple paths from ${\cal L}_{v}$ will not add the same safe path to ${\cal L}_{v^*}$ as it will start from a leaf only when the last such path is processed. This also avoids adding a non-maximal path which is a subpath of another safe path. 
\sbzk{CASE WISE DESCRIPTION???}

Finally, we need to add the safe paths to non-preferred maximum outgoing neighbours, which by \Cref{prop:nonMaxOut} is always on a single path containing the unique maximum incoming edges. We thus compute the single safe left extension for all such neighbours explicitly using \Cref{thm:flowPropO}\ref{prop:addEdge}. We do this again by maintaining the excess flow of the path from $x$ to $v$ in $f_x$. As we start the path is a single edge with $f_x$ necessarily positive, where we continue adding the preferred maximum incoming edge to the left until $f_x$ is negative. Note that we do not insist on a unique maximum incoming edge as required by \Cref{prop:nonMaxOut} as the flow $f_x$ will itself become negative if the preferred maximum incoming edge is not unique.

\begin{algorithm}[tbh]
	\caption{Optimally Computing Raw representation of Safe Paths}
	\label{alg:optSimp}

	\DontPrintSemicolon
	\BlankLine

	\begin{multicols}{2}
	Compute Topological Order of $G$\;
    \ForAll{$u\in V$ in topological order}{
        \textsc{Compute-Safe}$(u)$\;
    }
    \BlankLine
\!\!\!\textsc{Compute-Safe$(u)$:}
    
   \uIf{$u\notin Sink(G)\cup Source(G)$}
        {$v^*\gets v_{max}(u,\cdot)$}
    \lElse{$v^*\gets null$}
 
    \lIf{$u\in Source(G)$}
        {Initialize ${\cal T}_u$ with $u$}
    \BlankLine
    
    \ForAll(\tcp*[f]{new paths}){$(u,v)\in G, v\neq v^*$}{
            Add $(u,v)$ to ${\cal T}_v$\; 
            $x\gets u$, $f_x\gets f(u,v)$\;
            
            \While{$x\notin Source(G)$ \textbf{and} $f_x-f_{in}(x)+f_{max}(\cdot,x)>0$}{
                Add $e_{max}(\cdot,x)$ to ${\cal T}_v$\;
                $f_x\gets f_x-f_{in}(x)+f_{max}(\cdot,x)$\;
                $x\gets v_{max}(\cdot,x)$\; 
            }
            Add $(x,f_x)$ to ${\cal L}_v$\;
        }
 
     \If{$v^*\neq null$}{
        Make ${\cal T}_u$ as child of $v^*$ in ${\cal T}_{v^*}$\;        
            }

    \ForAll(\tcp*[f]{Process ${\cal T}_u$}){$(x,f_x)\in {\cal L}_u$}{
        \If{$v^*=null$ \textbf{or} $f_x-f_{out}(u)+f(u,v^*)\leq 0$}{
                        $p\gets$ Extract path from $x$ to $u$ in ${\cal T}_u$\;
                Add $(p,f_x)$ to \textsc{Sol}\;}
                
        \If{$v^*\neq null$}{
            $f_x\gets f_x-f_{out}(u)+f(u,v^*)$\;
            \While{$f_x\leq 0$ \textbf{and} $x$ is leaf of ${\cal T}_{v^*}$}{
                    $y\gets $ Parent of $x$ in ${\cal T}_{v^*}$\;
                    $f_x\gets f_x+f_{in}(y)-f(x,y)$\;
                    Remove $(x,y)$ from ${\cal T}_{v^*}$\;
                    $x\gets y$\;
                }
                
                \If{$x$ is a leaf in ${\cal T}_{v^*}$}{     Add $(x,f_x)$ to ${\cal L}_{v^*}$\;
                }
            
        }            
        }
    
   \end{multicols}
\end{algorithm}
\subsection{Correctness}
We prove the correctness of the algorithm by induction over the topological order of the graph. The underlying invariant is as follows:

\textit{After \textsc{Compute-Safe}$(u)$ is executed, all {\em left maximal} safe paths having starting vertex and the internal vertices with topological order up to $u$, are stored in corresponding ${\cal T}_v$ and ${\cal L}_v$. Also, all {\em maximal} safe paths ending at vertices with topological order up to $u$, are reported in \textsc{Sol}.} 

The base case is trivially true when no vertices are processed, as no safe paths exist. Now, when we start processing $u$, using the invariant we know all the {\em left maximal} safe paths not having $u$ as internal vertex, and all the {\em maximal} safe paths ending at the vertex with topological order less than $u$ are already in \textsc{Sol}. So we need to process only the {\em left maximal} paths having $u$, which necessarily have the last internal vertex as $u$, and all the {\em maximal} safe paths ending at $u$ must be added to \textsc{Sol}. The prefix (not necessarily proper) of both these kinds of paths up to $u$, are clearly safe (using \Cref{thm:flowPropO}\ref{prop:addEdge}) and hence are present in ${\cal T}_u$ and ${\cal L}_u$ by the invariant. 

Now, all the {\em left maximal} safe paths are checked for a possible extension to $v^*$ by construction, and for the remaining out-neighbours we explicitly add the single safe path possible (\Cref{prop:nonMaxOut}). Further, all the paths in ${\cal L}_u$ are checked for being maximal and added to \textsc{Sol} in such a case. Note that processing the vertices in the topological order ensures that all safe paths ending at $u$ have internal vertices already processed so that the complete path is present in ${\cal T}_u$ and ${\cal L}_u$ before it is processed. 

\subsection{Analysis }
The total time required by the algorithm can be associated with the edges of the graph $m$ or the total length of safe paths reported, i.e. size of the raw representation of the output $out_R$. Computing the topological order of the graph requires $O(m)$ time. Now, for each vertex $u$, processing the paths in ${\cal L}_u$ either extends it to $v^*$ in $O(1)$ time or reporting a safe path $p$ and extending its subpath to $v^*$ in $O(|p|)$ time. In the former case, the length of the safe path is increased by one (adding $u$), and the latter case is associated with the length of the reported path (as each safe path is reported exactly once). For residual out-neighbours, the time required is proportional to the size of added safe path. Hence we have the following: 

\begin{restatable}{theorem}{optCompSafePath}
Given a flow graph (DAG)  having $n$ vertices and $m$ edges, the set of all safe paths can be optimally reported in its raw representation in $O(m+out_R)$ time.
\end{restatable}

\section{Optimal computation of the optimal Concise Representation}
\label{sec:optOCompR}

Previous work~\cite{KhanMMLA22} presented a simple algorithm for computing the concise representation of the solution. Hence instead of reporting each safe path individually which may have overlaps among each other, they combined several overlapping safe paths to report a single path $p$ along with indices representing the subpaths of $p$ which are maximally safe.

However, 
their concise representation was dependent on the underlying candidate path decomposition, which may be suboptimal. 
We shall now present an optimal algorithm for computing the optimal concise representation of the solution. Our algorithm again uses Path Tries ${\cal T}_u$ with a modified version ${\cal L}^+_u$ of the list of safe paths to store the concise representation. 

\subsection{Data structures ${\cal T}_u$ and ${\cal L}^+_u$}
We again build a Trie structure on the reverse paths, whose common suffixes are combined. Similar to the previous algorithm, it stores all the left maximal safe paths ending at $u$. 

Now, the concise representation of safe paths are reported in the form $\{(p_1,I_1),(p_2,I_2),\cdots\}$, where each path $p_i$ has maximal safe subpaths denoted by intervals $I_i=\{(l_1,r_1,f_1),(l_2,r_2,f_2),\cdots\}$. Each $(l_j,r_j)$ and $f_j$ denote the corresponding end vertices of the maximal safe path on $p_i$ and its excess flow, respectively. 
While processing $u$, the partial results are maintained in the list ${\cal L}^+_u =\{(p_1,I_1),(p_2,I_2),\cdots\}$ where $p_i$ are the partially built concise representation of the safe paths, where the reported paths contain $u$. Further, for each $I_i$ the last interval $(l_j,r_j,f_j)$  has $r_j=u$ representing the left maximal safe path ending at $u$, whereas the remaining intervals are maximal. Note that while processing $u$, $p_i$ does not include the last path from $l_{j}$ to $u$.

\subsection{Algorithm}
The main idea behind our approach is to always attempt to extend a path $p_i$ of the concise representation with the interval corresponding to a safe path that overlaps the most with $p_i$. Clearly, while extending the left maximal safe paths on $u$ to its out-neighbours, the maximum such overlap with $p_i$ would correspond to the safe path ending at the preferred maximum outgoing neighbour $v^*$, which is hence added to $p_i$. However, in case multiple paths $p_i,p_j$ in the concise representation add exactly the same safe path $p_{v^*}$ corresponding to $v^*$, it is not optimal to add $p_{v^*}$ to both $p_i$ and $p_j$. In such a case $p_{v^*}$ can be added to anyone such path (say $p_i$), and $p_j$ will add the maximum overlapping path $p_v$ corresponding to some other out-neighbour $v$ of $u$, if it exists. If no such neighbour exists, we will report $p_j$ in the solution having the last interval ending at $u$. And in case the safe path $p_{v'}$ of some out-neighbour $v'$ of $u$ is not accommodated in the existing paths of the concise representation, we add a new path $p_{v'}$ to the concise representation. The optimality of our concise representation is guaranteed by our choice of maximum overlap, ensuring a new path in the concise representation is always of the minimum length.

Consider \Cref{alg:optConR}, similar to the previous algorithm we process each vertex in the topological order, and when a vertex $u$ is processed its ${\cal T}_u$ and ${\cal L}^+_u$ have already been computed by its incoming neighbours. Similar to the previous approach for each out-neighbour $v$ of $u$, that is not the preferred maximum out-neighbour of $u$ exactly a single safe path exists containing the unique maximum incoming edges (\Cref{prop:nonMaxOut}). We compute it similar to the previous algorithm for each $v$ from scratch, and update its corresponding ${\cal T}_v$, inserting the path $p_v$ temporarily to ${\cal L}^+_v$ as a new path. This path can potentially be added to some existing path in ${\cal L}^+_u$, for which we mark the starting vertex of $p_v$ in ${\cal T}_u$.

Now, for the unique maximum outgoing neighbour $v^*$ of $u$, we add ${\cal T}_u$ as a child of $v^*$ in ${\cal T}_{v^*}$ and attempt to extend each path $p_k$ in ${\cal L}^+_u$ to $v^*$ using \Cref{thm:flowPropO}\ref{prop:addEdge}. Similar to the previous algorithm, this is accompanied by triming the leaves of last interval of $p_k$ from ${\cal T}_u$ if they are not safe in ${\cal T}_{v^*}$. Again, if the start of the safe path $x$ for $v^*$ is no longer a leaf, then the same safe subpath is shared by some other safe path in ${\cal L}^+_{u}$. In such a case we do not extend $p_k$ to $v^*$, rather either (a) extend it to the out-neighbour $v$ of $u$ having the maximum overlap (lowest vertex marked in ${\cal T}_u$ along the last interval of $p_k$) which is not a unique maximum out-neighbour of $u$, or (b) terminate $p_k$ at $u$ including it as its last interval. Thus, while processing ${\cal T}_u$ for each path in ${\cal L}^+_u$ we deal with five distinct cases (see \Cref{alg:optConR}).

\begin{enumerate}[(a)]
    \item \textbf{$v^*$ is null because $u\in Source(G)$:} The list ${\cal L}^+_u$ is empty and each out-neighbour $v$ of $u$ is addressed as non-preferred maximum out-neighbour adding the corresponding edge to their ${\cal T}_v$ and ${\cal L}^+_v$ computing from scratch.
    \item \textbf{$v^*$ is null because $u\in Sink(G)$:} For all paths in ${\cal L}^+_u$, the left limit $x$ reaches $u$ (as $f_x$ is always negative and no out-neighbour exists to mark a vertex in ${\cal T}_u$), which is hence updated to include the last interval of $I_k$ and added to \textsc{Sol}.
    \item \textbf{Path $p_k$ is extended to include $v^*$:} The safe path is unique to $p_k$ and hence the left limit of safe path $x$ is always a leaf, terminating as soon as $f_x>0$ and added to ${\cal L}^+_{v^*}$ accordingly. Note that no vertex before reaching $f_x>0$ could have been marked, as left limit of $v^*$ would be the lowest among all out-neighbours of $u$.
    \item \textbf{Path $p_k$ is extended to include some $v\neq v^*$:} This is possible only if the safe path for $v^*$ is not unique to $p_k$, i.e., $x$ is no longer a leaf. Then maximum overlap is the lowest vertex along $x$ to $u$ path, to which $p_k$ is added accordingly.
    \item \textbf{Path $p_k$ is not extended and reported in \textsc{Sol}:} This is possible again when safe path for $v^*$ is not unique, so $x$ is no longer a leaf and $x$ reaches $u$ similar to case (b).
\end{enumerate}

\subsection{Correctness and Analysis}
The optimality of $out_C$ is ensured by appending a path in $out_C$ with the safe path having the maximum overlap with the existing path. This is ensured by processing the marked vertices bottom-up, which represent the start vertex of the safe paths corresponding to the out-neighbour $v$ of $u$, which is not a unique maximum out-neighbour of $u$. This guarantees that in case the path cannot be uniquely extended to $v^*$, the vertex $v$ with the maximum overlap is selected resulting in an optimal concise representation.

The total time taken while processing $u$ is dominated by the processing of ${\cal L}^+_u$ and building the safe paths for those out-neighbours of $u$ which are not unique maximum out-neighbours of $u$, from scratch. Consider the cases (a), (b), (c) and (e), the time taken in processing ${\cal L}^+_u$ can be easily associated with the length of the path $p_k$ in ${\cal L}^+_u$ since it will be reported exactly once (cases (b) and (e)), and removed from the last interval exactly once (case (c)). Case (a) is also easy to associate as it increases the corresponding paths in ${\cal L}^+_v$, and hence can be associated with its length. 

The only hard case is (d) as it computes the safe path for $v$ from scratch, but extends it on an existing $p_k\in {\cal L}^+_u$ which takes more time than the increase in $p_k$. However, note that this is possible only when the left limit $x$ is no longer a leaf, which implies that the extra processed path is a common suffix of multiple paths in ${\cal L}^+_u$. And hence the suffix was accounted for only once for multiple paths, and now when $p_k$ is detached from the common suffix the cost of the processed path can be associated with that of the detached path (previously unaccounted being a part of the common suffix). Thus, all the steps can be accounted for with the length of $out_C$ and we get the following.

\begin{restatable}{theorem}{optCR}
Given a flow graph (DAG)  having $n$ vertices and $m$ edges, the optimal concise representation $out_C$ of the safe paths can be optimally reported in $O(m+out_C)$ time.
\end{restatable}

\begin{algorithm}[tbh]
	\caption{Optimally Computing concise Representation of Safe Paths}
	\label{alg:optConR}

	\DontPrintSemicolon
	\BlankLine

	\begin{multicols}{2}
	Compute Topological Order of $G$\;
    \ForAll{$u\in V$ in topological order}{
        \textsc{Compute-Safe-CompR}$(u)$\;
    }
    \BlankLine
    \BlankLine
    \BlankLine
    \BlankLine
\!\!\!\textsc{Compute-Safe-CompR$(u)$:}
    \uIf{$u\notin Sink(G)\cup Source(G)$}
        {$v^*\gets v_{max}(u,\cdot)$}
    \lElse{$v^*\gets null$}

    \lIf{$u\in Source(G)$}{
    Initialize ${\cal T}_u$ with $u$}
     
    \BlankLine
       
        \ForAll(\tcp*[f]{new paths}){$(u,v)\in G, v\neq v^*$}{
            Add $(u,v)$ to ${\cal T}_v$\;
            $x\gets u$ in ${\cal T}_u$, $f_x\gets f(u,v)$\;
            
            \While{$x\neq$ leaf of ${\cal T}_u$ \textbf{and} $f_x-f_{in}(x)+f_{max}(\cdot,x)>0$}{
                Add $e_{max}(\cdot,x)$ to ${\cal T}_v$\;
                $f_x\gets f_x-f_{in}(x)+f_{max}(\cdot,x)$\;
                $x\gets v_{max}(\cdot,x)$ in ${\cal T}_u$\; 
            }
           
            Add $(\emptyset,\{(x,v,f_x)\})$ to ${\cal L}^+_v$\;
            Push $v$ to $Mark[x]$\;
            Add $x$ to $\cal M$\;
        }
 
     \If{$v^*\neq null$}{
        Add ${\cal T}_u$ as child of $v^*$ in ${\cal T}_{v^*}$\;        
            }

    \ForAll(\tcp*[f]{Process ${\cal T}_u$}){$(p_k,I_k)\in {\cal L}^+_u$}{
        $(l_i,u,f_i)\gets$ Last of $I_k$, 
        $x\gets l_i$ in ${\cal T}_u$\;
        \lIf{$v^*= null$}{$f_x\gets -\infty$}
        \lElse {$f_x\gets f_i-f_{out}(u)+f(u,v^*)$}
        \BlankLine        

         \While{$f_x\leq 0$ \textbf{and} $Mark[x]=\emptyset$ \textbf{and} $x\neq u$}{
                    $y\gets $ Parent of $x$ in ${\cal T}_{u}$\;
                    \If{$y$ is not a leaf in ${\cal T}_{u}$}{
                    $f_x\gets f_x+f_{in}(y)-f(x,y)$\;
                    Remove $(x,y)$ from ${\cal T}_{u}$\;
                    }
                    $x\gets y$\;
                }
                
                $p\gets$ Path from $l_i$ to $x$ in ${\cal T}_{u}$\; 
                $p_k\gets p_k\cup \{p\setminus \{x\}\}$\;  
                   
                \uIf{$f_x>0$}{
                \lIf{$l_i\neq x$}{
                    Add $(x,v^*,f_x)$ to $I_k$
                    }\lElse{
                    Last of $I_k\gets (l_i,v^*,f_x)$
                    }    
                    Add $(p_k,I_k)$ to ${\cal L}^+_{v^*}$\;
                }
                \Else{
                    \uIf{Mark$[x]\neq \emptyset$}{
                    $v\gets $ Pop from Mark$[x]$\;
                    $(\emptyset,I_v)\gets $ Pop from ${\cal L}^+_v$\;
                    Add $(p_k,I_v\cup I_k)$ to ${\cal L}^+_v$\;
                    }\lElse{
                    Add $(p_k\cup \{x\},I_k)$ to \textsc{Sol}
                    }
                }
        }            
    \lForAll{$x\in {\cal M}$}{Clear Mark$[x]$}       
    Clear ${\cal M}$\;
    
   \end{multicols}
\end{algorithm}

\section{Optimal Representation of Safe paths}
The raw representation of the safe paths $out_R$ can take $\Theta(mn^2)$ space and hence time in the worst case. The previous work~\cite{KhanMMLA22} presented a concise representation of the safe paths reporting a combination of the overlapping safe paths along with its indices, requiring total $\Theta(mn)$ space. However, it may not be optimal as the total size of this concise representation  may be much larger than the number of safe paths. We thus present an optimal representation of the safe path whose size requires $\Theta(1)$ space for every safe path reported. 

\subsection{Representative edge with left and right extensions}
\Cref{prop:nonMaxOut} presents an interesting property about safe paths being extendible in a preferred way to the left for edges which are not unique maximum outgoing edges of some vertex. We extend the notion further by considering a representative edge for each safe path such that the maximal path can always be generated by extending it to the left along unique maximum incoming edges and to the right along unique maximum outgoing edges as follows.

\begin{theorem}[Representative edge]
Given a flow graph (DAG), every safe path $p$ can be described using a representative edge $e_p$, such that $p$ can be constructed by extending $e_p$ to the left along the unique maximum incoming edges and to the right along the unique maximum outgoing edges. 
\label{thm:repE}
\end{theorem}
\begin{proof}
Given a maximal safe path $p$, let $e(x,y)\in p$ be the leftmost edge such that $e(x,y)\neq e^*_{max}(\cdot,y)$. If no such edge exists, we define the last edge as the representative edge of $p$, where rest of $p$ is along its unique maximum incoming edges (left extension of $e$) proving the existence of the representative edge.

Now, by definition the prefix of $p$ on the left of $e$ is along the unique maximum incoming edges (left extension of $e$), so we only need to prove that the suffix of $p$ after $e$ is along the unique maximum outgoing edges (or the right extension of $e$). We shall prove it by contradiction, hence assume there exist an edge $e'(a,b)\in p$ to the right of $e$ such that $e'(a,b) \neq e^*_{max}(a,\cdot)$. Clearly, the path $e^*_{max}(\cdot,y)\cup p[y,a]\cup e^*_{max}(a,\cdot)$ is also safe using \Cref{thm:flowPropO}\ref{prop:addEdge}. However, this violates the merge-diverge property (\Cref{prop:merge-diverge}) contradicting our assumption and proving the existence of $e$ as the representative edge of $p$.
\end{proof}

\begin{remark}
Every safe path contains a representative edge which is either the last edge which is also unique maximum incoming edge, or an edge which is not a unique maximum incoming edge. For the sake of uniformity, in case multiple edges satisfy this property (not being unique maximum incoming edge) for a safe path $p$, we consider $e_p$ to be the rightmost such edge. 
\label{rem:repEdgeType}
\end{remark}

Note that the safe paths when represented using such a representation requires $O(1)$ space per safe path to store the representative edge and its two endpoints, where a single representative edge may store multiple pairs of endpoints representing individual safe paths. We assume that the unique maximum incoming and outgoing edges of each vertex are known which can be pre-computed in $O(m)$ time and stored using $O(n)$ space.

\subsection{Approach}
As described previously in \Cref{rem:repEdgeType}, the representative edge $e$ can be either (a) an edge which is  not aunique maximum incoming edge, or (b) a unique maximum incoming edge. The former case is non-trivial as the edge can represent multiple safe paths, whereas the latter is trivial as the edge can represent at most one (can be zero) maximal safe path ending at $e$. So here we describe only the non-trivial case as trivial can be computed similarly. 

\begin{figure}
    \centering
    \includegraphics[scale=.8]{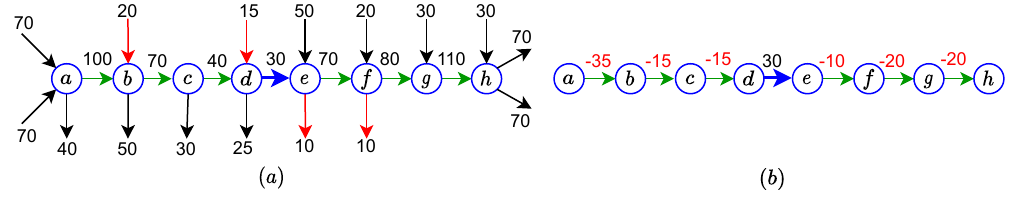}
    \caption{Graphs describing (a) the left and right extensions of a representative edge, and (b) the cumulative losses on extension along each edge. }
    \label{fig:approach}
\end{figure}

Consider \Cref{fig:approach} (a), where the edge $(d,e)$ is a edge which is not a unique maximum incoming edge of $e$. The path $<a,b,c,d>$ is its left extension (along unique maximum incoming edges), and the path $<e,f,g,h>$ is its right extension (along unique maximum outgoing edges). Now, once we compute the left and right extensions we can easily compute all the safe paths represented by $(d,e)$ using two pointer approach in time proportional to length of the path. We get the maximal safe paths $<b,c,d,e,f>$ and  $<d,e,f,g,h>$.

However, assuming we have pre-computed the left and right extensions we can compute all the safe paths using binary search along the path in time $O(\log n)$ times the number of safe paths. This can be done by pre-computing the loss for extension along each edge (\Cref{thm:flowPropO}\ref{prop:addEdge}) and storing the cumulative value on the edge. Now, we find the safe paths as follows. Given the flow on $(d,e)$ is $30$ we search for the leftmost minimum value greater than $-30$ which we find as $(b,c)$ with value $-15$ giving a left maximal safe path. We make it right maximal (and hence maximal) by searching for the rightmost minimum value greater than $-30+15=-15$ which we find as $(e,f)$, giving our maximal safe path from $b$ to $f$ using two binary searches. Now, to find the next maximal we extend it to right including $(f,g)$ and again find the left maximal on $-30+20$ as $(d,e)$, and thereafter right maximal on $-30+0$ as $(g,h)$, giving the second maximal safe path from $d$ to $h$ using another two binary searches.

\subsection{Data structures}
As described in our approach above we require pre-computed left and right extensions for each edge along with the cumulative losses on each edge for efficient search of the maximal safe paths. This can be efficiently computed by building all possible left and right extensions separately which will form two forests corresponding to all unique maximum incoming and outgoing edges. Further for $O(1)$ time access to elements on these forests for binary search we require the classical Level ancestor data structure.

\begin{enumerate}
    \item \textbf{Unique maximum incoming and outgoing forests ${\cal F}_i$ and ${\cal F}_o$.}\\
    For each vertex in the graph, we add the unique maximum incoming edge (if exists) to ${\cal F}_i$. Clearly, each vertex has at most one incoming neighbour (parent) making ${\cal F}_i$ a forest. Similarly, for each vertex, we add the {\em reverse} of the unique maximum outgoing edge (if exists) and ${\cal F}_o$. Again, each vertex has at most one incoming neighbour as a parent (as we added reverse edges), making ${\cal F}_o$. 
    
    Now, for each vertex $v$ in the forest ${\cal F}_i$ we store the cumulative loss $c_i[v]$ (using \Cref{thm:flowPropO}\ref{prop:addEdge}) along the path $v$ to the root of its tree $root[{\cal F}_i(v)]$. The cumulative loss for any subpath from $v$ to $u$ (where $u$ is ancestor of $v$ in ${\cal F}_i$), can be simply computed as $c_i[u]-c_i[v]$. Similarly, we store the cumulative loss for each vertex $v$ on ${\cal F}_o$ in $c_o[v]$. The corresponding forests ${\cal F}_i$ and ${\cal F}_o$ for \Cref{fig:baseGraph} are shown in \Cref{fig:forests}. Clearly, these structures can be computed in $O(m)$ time.

\begin{figure}
    \centering
    \includegraphics[scale=.8]{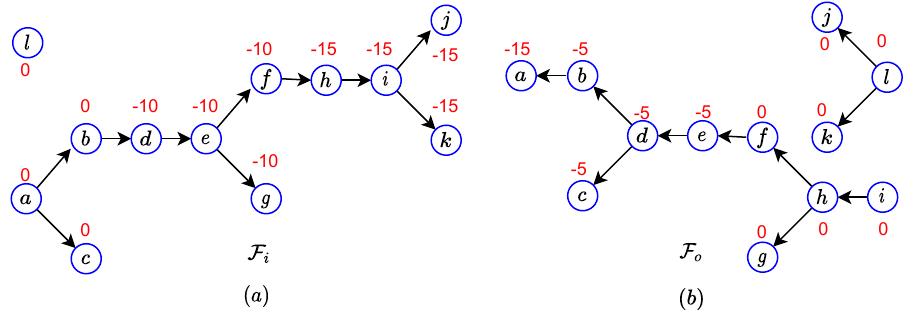}
    \caption{For the graph in \Cref{fig:baseGraph} we show (a) the unique maximum incoming forest ${\cal F}_i$ and (b) the unique maximum outgoing forest ${\cal F}_o$. Note that $l$ (in ${\cal F}_i$) and $i$ (in ${\cal F}_o$) do not have a parent in the absence of corresponding unique maximum edges.}
    \label{fig:forests}
\end{figure}

    \item \textbf{Level ancestors $LA_i$ and $LA_o$}\\
    We use \Cref{thm:LA} to compute the data structure on ${\cal F}_i$ and ${\cal F}_o$ using $O(n)$ time,  for reporting the level ancestors $LA_i(v,d)$ and $LA_i(v,d)$ of a vertex $v$ at depth $d$ in $O(1)$ time.
\end{enumerate}

\subsection{Algorithm}
We now describe how our approach (described on paths) can be used to compute the optimal representation of all the maximal safe paths in $O(m+out_O\log n)$ time.  

We first address computing all non-trivial safe paths. Consider each edge which is not unique maximum incoming edge, say $e(u,v)\neq e^*_{max}(\cdot,u)$. We use the two pointer algorithm as described in~\cite{KhanMMLA22} however in each step we perform a binary search to compute each safe path in $O(\log n)$ time. The binary search computes the excess flow on a path using the  values of $c_o[v], c_i[v]$, and probes an element by considering ancestors of $u$ in ${\cal F}_i$ and ancestors of $v$ in ${\cal F}_o$ which can be directly accessed in $O(1)$ time using $LA_i$ and $LA_o$ structures. The optimal output reports a list of pairs of end vertices for maximal safe paths represented by $e(u,v)$. However, we avoid this algorithm if the maximal safe path containing $e$ is the single edge $e$, which is typically not reported. This can be evaluated in $O(1)$ time using \Cref{thm:flowPropO}\ref{prop:addEdge} in both left and right directions.

For computing the trivial paths, we need to compute all maximal safe paths containing only unique maximum incoming edges. We simply look at the leaves of ${\cal F}_i$ and search for the left maximal path ending on it, and thereafter continue the search using a modified two pointer algorithm (using binary search) on its ancestors. However, we need to ensure two aspects. Firstly, we do not perform a binary search in case there exist no maximal safe path, which is only possible if (a)  $f(u,v)>|c_i[u]|$, implying the entire path to root is safe, and (b) $f(e_{max}(v,\cdot))>|c_i[v]|$ implying the path till $v$ is not maximal. In case only (a) is true we report a single safe path from the root to $v$, and if both are true we skip the leaf. Secondly, the two pointer algorithm on internal vertices may repeat the safe paths reported by other leaves as the internal vertices may be shared. Hence we mark the internal vertices which have been processed so as to avoid repetition of processing and results. 

\subsection{Implementation details}
For the sake of completeness we now describe the two pointer algorithm using binary search on ${\cal F}_i$ and ${\cal F}_o$ in detail. 

The algorithm computes all pairs of end points for safe paths represented by $e(u,v)$ as follows. It first computes the start of left maximal safe path ending at $v$ by performing a binary search on the ancestors of $u$ in ${\cal F}_i$. It computes the highest ancestor $l$ of $u$ such that excess flow of $l$ to $v$ is positive, i.e. $c_i[l]< c_i[u]+f(e)$. This search involves accessing the ancestor at mid-depth directly in $O(1)$ time using $LA_i(u,d[u]/2)$ and so on for the whole binary search. Thereafter, the algorithm computes the end $r$ of right maximal path starting from $l$ in ancestors of $v$ in ${\cal F}_o$ such that excess flow of the path $<l,\cdots,u,v,\cdots,r>$ is positive, i.e.  highest ancestor of $v$ such that to $c_o[r]<f(e)+c_i[u]+c_o[v]-c_i[l]$. We record $[r,l]$ with the corresponding flow $f(e)+c_i[u]-c_i[l]+c_o[v]-c_o[r]$ as a safe path for $e$. Then we find the next start of the left maximal path ending at ancestor of $r$ in ${\cal F}_o$, i.e., using  $f(e)+c_o[v]-c_o[r]$ instead of $f(e)$ in the process described above. This continues until the left maximal reaches $v$ or the right maximal reaches the root of ${\cal F}_o$ containing $v$.

\subsection{Correctness and Analysis}
The correctness of our algorithm follows from that of the two pointer algorithm described in~\cite{KhanMMLA22}. By construction, we show that the algorithm requires $O(1)$ time to check whether a safe path exists corresponding to a representative edge, and thereafter $O(\log n)$ time to report each safe path represented by the edge. Since any explicit representation of the safe paths would require $O(1)$ space for every safe path, $out_O$ is the number of safe paths. We thus get the following result.

\begin{restatable}{theorem}{optR}
Given a flow graph (DAG)  having $n$ vertices and $m$ edges, the optimal representation $out_O$ of the safe paths can be reported in $O(m+out_O\log n)$ time.
\end{restatable}

\section{Space bounds for different representations of safe paths}
Previous work~\cite{KhanMMLA22} presented a worst case example demonstrating that $out_R$ and $out_C$ can respectively be $\Omega(mn^2)$ and $\Omega(mn)$ in the worst case and $O(m)$ in the best case. The worst case example graph they presented also gives a bound of $\Omega(mn)$ in the worst case and $O(m)$ in the best case for $out_O$. In the light of the new characterization, we shall now understand these bounds in more detail.

Using \Cref{thm:repE}, we know that every safe path can be represented as an edge with a subpath of its left and right extensions. Now, in the worst case each of the edge $m$ edges can have left and right extensions of length $O(n)$ each, making a complete path of $O(n)$ size. Using the two pointer algorithm we know, that the number of maximal safe paths on a path $p$ are $|p|$. Hence, for each edge we can have $O(n)$ safe paths, each of possibly $O(n)$ edges in the worst case. 

This establishes an upper bound of $O(mn^2)$ on the size of $out_R$ matching the $\Omega(mn^2)$ bound of \cite{KhanMMLA22}, proving the tight bound of $\Theta(mn^2)$ on $out_R$ in the worst case. Further, since each edge with its extensions creates a valid path for $out_C$, having $O(n)$ length and $O(n)$ indices on every path, we also get $O(mn)$ bound for $out_C$ and $out_O$, resulting in tight worst case bound of $\Theta(mn)$ for both $out_C$ and $out_O$.

\newpage

\section{Conclusion}
We study the optimization of the solutions for the safety of flow paths in a given flow graph (DAG), which has applications in various domains, including the more prominent assembly of biological sequences. The previous work characterized such paths giving an optimal verification algorithm but suboptimal enumeration algorithms, which required computing a candidate flow decomposition taking $\Omega(mn)$ time even when the reported solution is small. 

We present output-sensitive optimal algorithms for reporting the safe paths when represented in the raw  format reporting each path explicitly, and optimal concise representation previously described. This is achieved by exploiting a  crucial property related to the interaction of safe paths and a novel data structure Path Tries, which may be of independent interest. Further, we characterized an optimal representation of the safe paths, requiring $O(1)$ space for every safe path reported. We also presented a near optimal algorithm to compute the optimal representation of the safe paths.  The new characterization additionally allows us to understand the space bounds of various representations of all safe paths, where we match the existing lower bounds with worst case upper bounds.

In the future, it would be interesting to see an optimal output-sensitive algorithm for computing even the optimal representation of the safe paths  (dropping the $O(\log n)$ factor). It would also be interesting to see if similar properties or algorithms can be used to solve related problems as path covers, or path decomposition for general graphs.

\bibliography{main}


\end{document}